# Graphane: a two-dimensional hydrocarbon

Jorge O. Sofo,[1,2] Ajay S. Chaudhari,[1,2] and Greg D. Barber[2]

[1] Department of Physics, [2] Materials Research Institute,

The Pennsylvania State University, University Park, Pennsylvania, 16802, USA

## Abstract

We predict the stability of a new *extended two-dimensional hydrocarbon* on the basis of first-principles total energy calculations. The compound that we call graphane is a fully saturated hydrocarbon derived from a single graphene sheet with formula CH. All of the carbon atoms are in $sp^3$ hybridization forming a hexagonal network and the hydrogen atoms are bonded to carbon on both sides of the plane in an alternating manner. Graphane is predicted to be stable with a binding energy comparable to other hydrocarbons such as benzene, cyclohexane, and polyethylene. We discuss possible routes for synthesizing graphane and potential applications as a hydrogen storage material and in two dimensional electronics.

The prediction and discovery of new materials is always exciting for the promise of new applications and properties.[1, 2] It is even more fascinating when the material has some peculiar topological property such as reduced dimensionality and it is composed of just carbon and hydrogen. The principal theme of this report is the prediction of a new material, that we call graphane, which has all these properties and the eminent possibility of straightforward synthesis.

Hydrocarbons are the simplest organic compounds made only of carbon and hydrogen atoms.[3] Some of them occur naturally in the form of crude oil and natural gas. Others are synthesized such as polyethylene and other plastics. They are readily oxidized to produce carbon dioxide and water with a considerable release of energy; therefore, they are usually good fuels. All known hydrocarbons, until now, are molecules that consist of a carbon backbone with hydrogen atoms attached. The backbone can be a linear chain, a ring, or combinations of both. On the contrary, graphane, the compound we predict in this work, is the *first extended two-dimensional covalently bonded hydrocarbon*. Its fully dehydrogenated counterpart, graphene, has been the subject of many recent investigations due to its peculiar transport properties.[4-7] We predict that graphane is a semiconductor and, because of its novel structure and low dimensionality, it provides a fertile playground for fundamental science and technological applications. There is a fully fluorinated analog, poly-(carbon monofluoride) with formula CF, which has been synthesized before.[8] Because fluorine is known to replace hydrogen in many hydrocarbons, the existence of this fully fluorinated form gives further support to our prediction. Indeed, we will demonstrate that graphane is at least as stable as CF. In addition to the existence of the fluorinated analog, an isostructural compound with silicon replacing carbon was obtained by reaction of $CaSi_2$ with HCl. [9]

To motivate the experimental search for graphane, we present first-principles total energy calculations[10] to show that this compound has a very favorable formation energy, quite comparable with other hydrocarbons. We provide structural, electronic, and vibrational characteristics of this material and discuss possible pathways for synthesis; some of which are already being attempted.[11]

Our calculations were carried out using Density Functional Theory with a plane wave basis set as implemented in the CASTEP code.[10] The core electrons are treated with Vanderbilt ultrasoft pseudopotentials.[12] Exchange and correlation are treated within the generalized gradient approximation of Perdew, Burke, and Ernzerhof.[13] All calculations are done with a cutoff energy of 310 eV and a sampling of the Brillouin zone converged at least to the precision of the binding energies reported. These binding energies are calculated as the difference in total energy between the compound and a pseudoatomic calculation done for the same cutoff. The pseudoatomic total energies are -146.42 eV for carbon and -12.46 eV for hydrogen. The optimization of atomic positions and unit cell are stopped when the change in energy is less than $5 \times 10^{-6}$ eV/atom, the force on each atom is less than 0.01 eV/Å, the displacements are less than $5 \times 10^{-4}$ Å, and the stress on the cell is less than 0.02 GPa.

We find that graphane has two favorable conformations: a chair-like conformer with the hydrogen atoms alternating on both sides of the plane and a boat-like conformer with the hydrogen atoms alternating in pairs. A perspective view of the chair conformer is shown in Fig. 1. The space group, lattice parameter, atomic positions, and bond distances are given in Table 1. The unit cell of both conformers is displayed in the supplementary information provided online. In the chair conformer, every C-C bond connects carbon atoms with hydrogen attached at opposite sides of the plane. The calculated C-C bond length of 1.52 Å is similar to the $sp^3$ bond length of 1.53 Å in diamond and is much greater than 1.42 Å characteristic of $sp^2$ carbon in graphene. The boat conformer has two different types of C-C bonds: those connecting carbons bonded to hydrogen atoms on opposite sides of the plane with a bond length of 1.52 Å and those connecting carbon atoms bonded to hydrogen atoms on the same side of the plane with a bond length of 1.56 Å, slightly longer due to H-H repulsion. The C-H bond length of 1.1 Å is similar in both conformers and typical of hydrocarbon compounds.

**Table 1: Crystal structure and binding energies of the chair and boat conformations of graphane. The binding energy is the difference between the total energy of the isolated atoms and the compound.**

|  | Chair | Boat |
|---|---|---|
| Unit Cell | | |
|   Space group | P-3m1 (#164) | Pmmn (#59) |
|   A | 2.516 | 4.272 |
|   B | 2.516 | 2.505 |
|   C | 4.978 | 4.976 |
| Atomic positions | | |
|   C | (2d) (1/3, 2/3, 0.5419) | (4f) (0.8178, 0, 0.5636) |
|   H | (2d) (1/3, 2/3, 0.7479) | (4f) (0.7439, 0, 0.7717) |
| Bond length (Å) | | |
|   C-C | 1.52 | 1.56, 1.52 |
|   C-H | 1.11 | 1.10 |
| Binding energy (eV/atom) | 6.56 | 6.50 |

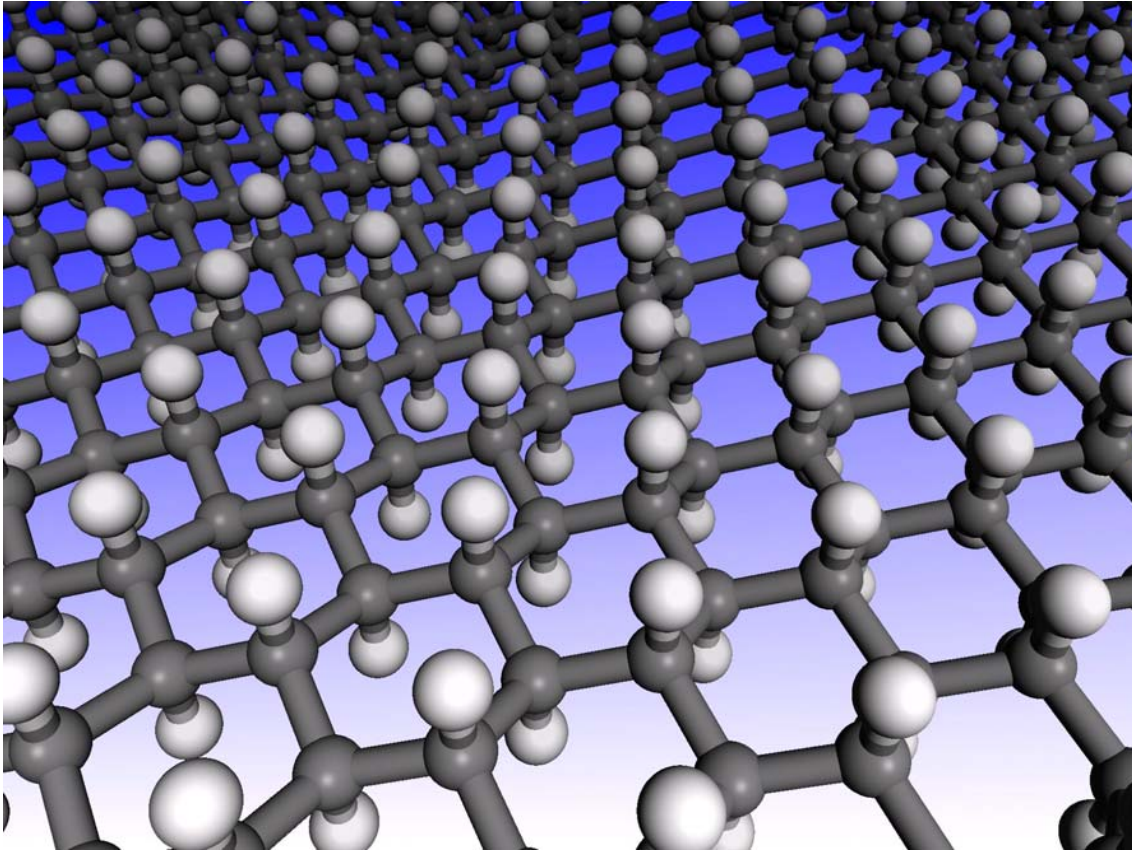

**Figure 1: Structure of graphane in the chair conformation. The carbon atoms are shown in gray and the hydrogen atoms in white. The figure shows the hexagonal network with carbon in the sp3 hybridization.**

The interplane bonding is weaker than between graphene planes. Physically, the separation between the planes reported here corresponds to their closest distance before the repulsion between their electronic clouds becomes important. We obtain this separation distance by minimizing the enthalpy under a uniaxial stress of 0.1 GPa, a small stress needed for the packing of graphane planes. Under this uniaxial stress, the graphane sheets are not bonded and the binding energy per atom is the same as in one sheet. The graphane bonds are fully saturated and there is no opportunity for hydrogen bonding between the sheets. The weak van der Waals attraction contributes negligibly to the results reported here.

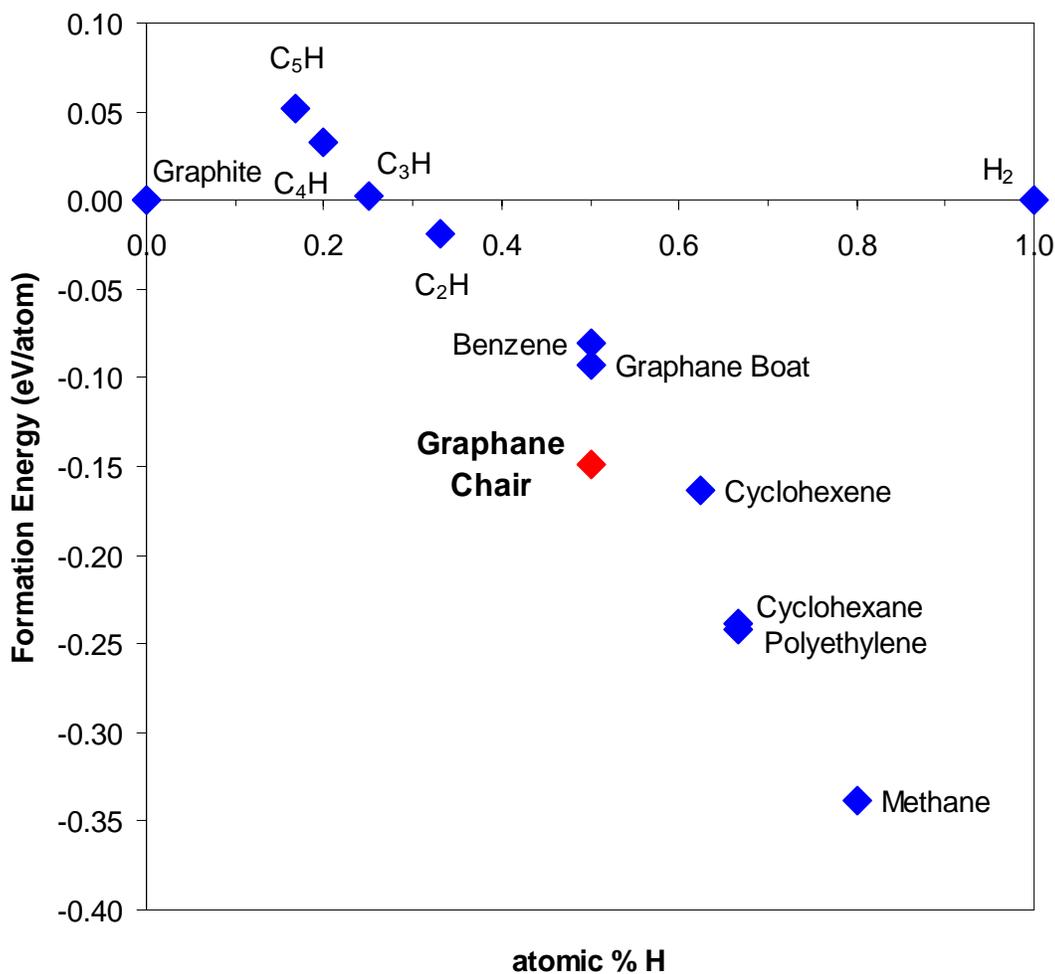

**Figure 2: Formation energy per atom as a function of hydrogen content in atomic percent. Graphane is among the most stable hydrocarbons and it is the most stable for its hydrogen concentration. It is also more stable than mixtures of cyclohexene and graphite.**

With a calculated binding energy of 6.56 eV/atom in the chair conformation, graphane is as stable as the analogous fluorinated compound CF. Using the same method of calculation we estimate the binding energy of CF to be 6.53 eV/atom. The binding energy of graphane in the chair conformation is even lower than the computed value for other hydrocarbons with 1:1 C:H ratio, such as benzene (6.49 eV/atom) and acetylene (5.90 eV/atom). Graphane is the most stable compound we found for this composition ratio. The binding energy difference between the chair and boat conformers of graphane is 0.055 eV/atom in favor of the chair. This energy difference is close to that in the fluorinated case, where similar electronic structure calculations obtained 0.0725 eV/atom also in favor of the chair conformer.[14] The boat configuration is not as stable as the chair conformation due to the repulsion of the two hydrogen atoms bonded to first neighbour carbon atoms on the same side of the sheet. As mentioned earlier, this repulsion stretches the C-C bond to 1.56 Å. A high binding energy is important and required for stability of the compound, but more important for the synthesis is the formation energy with respect to other competing hydrocarbons.

Stability with respect to other compounds is indicated by the formation energy. The natural references for the formation energy of this compound are graphite and $H_2$ at standard pressure and temperature conditions and for consistency we will use the theoretical values calculated with the same method. For graphite we will use our calculated binding energy of 9.55 eV/atom and in the case of $H_2$ we will use our calculated binding energy without zero point energy correction of 3.27 eV/atom. The formation energies per atom for different hydrocarbons are shown in Fig. 2 as a function of the atomic percent of H. In this plot, the point at (0,0) corresponds to graphite and the point at (1,0) corresponds to $H_2$. The figure shows that graphane fits in well with the family of already known and highly stable hydrocarbons such as polyethylene, benzene, cyclohexane, cyclohexene, and methane. Of the compounds with a C:H ratio of 1, graphane is the most stable and it is actually more stable than a mixture of graphite and cyclohexene with the same hydrogen concentration. This relative stability may explain the possible observation of graphane-like structures in ball milled samples of anthracite cyclohexene mixtures.[15] In Fig. 2 we have also included a series of compounds with formula $C_nH$ with n=2-5. The only member of this family with negative formation energy is $C_2H$. The chair conformer of graphane can be thought of as a monolayer of diamond (111) surface hydrogenated on both sides. $C_nH$ corresponds to a slab with n layers of carbon atoms cut from the same surface and hydrogenated on both sides. Other hydrogenated versions of existing fluorine graphite intercalation compounds show large and positive formation energy and thus should not be observed.

The electronic band structures of the two conformers are very similar. They have a direct band gap at the $\Gamma$ point with $E_g$=3.5 eV for the chair conformer and $E_g$=3.7 eV for the boat conformer. The band structure and corresponding density of states for the chair conformer are displayed in Fig. 3. The top of the valence band is doubly degenerate, with two different effective masses and decomposition of the density of states show that these states are mainly of p symmetry. On the contrary, the bottom of the conduction band exhibits mainly s character and has a larger effective mass. Graphane shows a moderate charge transfer from carbon to hydrogen (of the same order of magnitude as other hydrocarbons). A Mulliken population analysis[16] shows a transfer of about 0.2 electronic charges in both isomeric compounds.

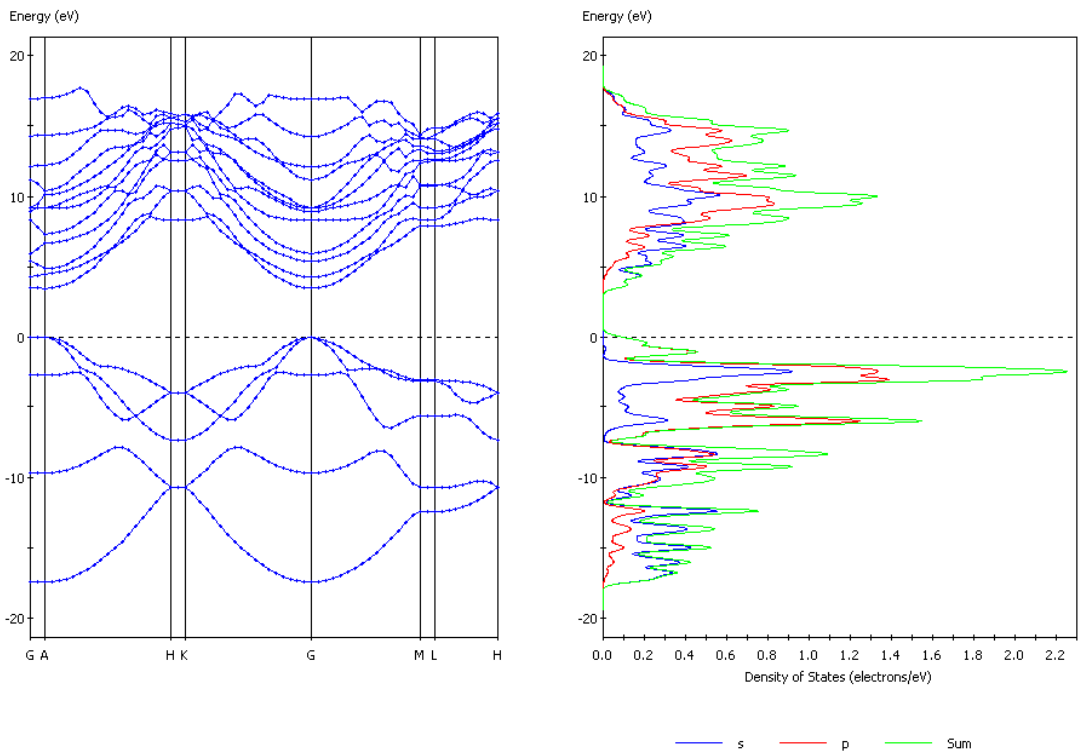

**Figure 3: Band structure (left) and density of states (right) of the chair conformer. The decomposition into states with s (blue) and p (red) symmetry is show together with the total density of states (green). The states at the top of the valence band are mainly of p symmetry while the states at the bottom of the conduction band have s symmetry.**

We have calculated the frequencies of phonons at the Γ point using a frozen phonon approach with a displacement of 0.0053 Å. The calculations are done for an isolated layer in a cell with a c lattice constant of 16 Å. The vibrational frequencies obtained are shown in Fig. 4. The highest frequency modes, corresponding to C-H bond stretching modes, occur at 3026 cm$^{-1}$ for the boat conformer and at 2919 cm$^{-1}$ for the chair. The frequency is higher for the boat conformer due to the interaction between neighboring hydrogen atoms on the same side of the plane. These C-H stretching modes are infrared active and should be useful in characterizing this compound.

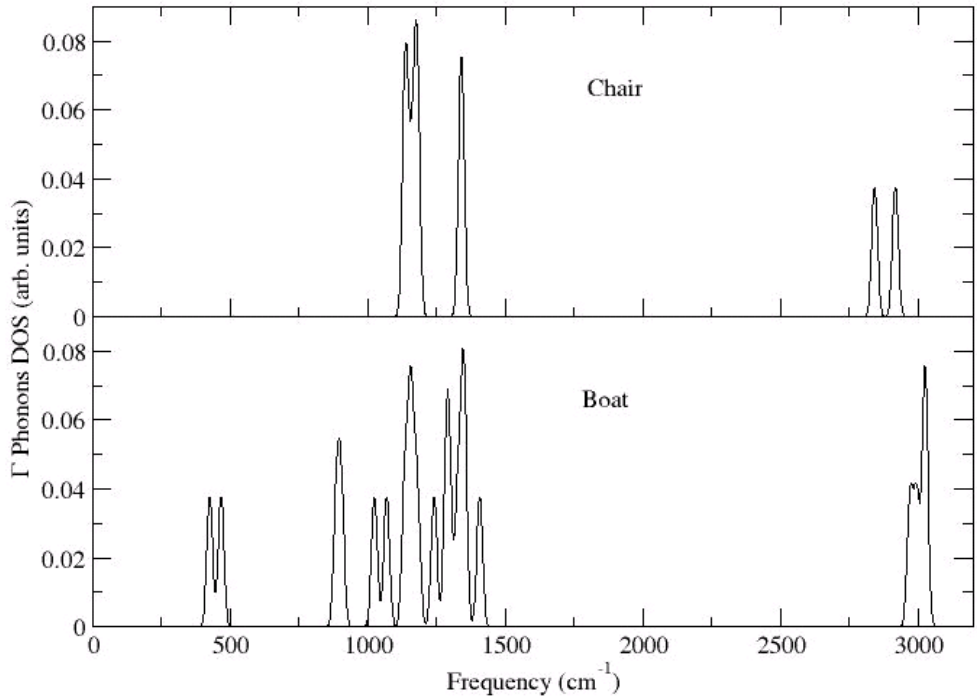

**Figure 4: Density of states of the phonons at Γ for the chair and boat conformers. The density of states is normalized to integrate to 3 times the number of atoms in the unit cell. The modes at higher frequency correspond to C-H bond stretching. In the boat configuration these modes are higher due to the lateral H-H interaction. The chair conformer is calculated in the hexagonal unit cell while the boat is calculated in the tetragonal unit cell with twice the number of atoms.**

We now turn our discussion to possible ways of synthesizing graphane. Hydrogenation of different forms of carbon has been thoroughly studied recently, including nanotubes,[17-20] different graphenic surfaces[21] and even bulk structures.[22] Adsorption of atomic hydrogen on graphite (0001) has been recently studied experimentally by STM[23] and theoretically[24] and on diamond (111).[25, 26] Direct exposure of graphite to an atmosphere of $F_2$ at high temperature produces fluorine intercalation compounds up to the monofluoride. However, direct exposure to $H_2$ does not seem to be the correct path to produce graphane because, unlike fluorine, hydrogen does not intercalate graphite[27] due to the higher binding energy of $H_2$ (~2.4eV/atom) compared to $F_2$ (~1.5eV/atom). Thus, the synthesis of graphane has to be directed through a different path. In the case of reactive ball milling between anthracite coal and cyclohexene, the delivery of H does not involve splitting molecular hydrogen. However, a method to produce cleaner samples would be desirable.

A possible synthetic path to graphane can start from CF and exchange fluorine by hydrogen. The direct exposure of $C_xF$ to hydrogen removes the fluorine from the compound to form HF and graphite, not graphane.[28] An alternative procedure being currently tested at Penn State for substituting fluorine with hydrogen is exchange with sodium hydride.[11] Our calculations show that the enthalpy change for this reaction is favorable by about 0.9 eV/atom. An alternative route explored in another context by Pekker et al. obtained layered hydrogenated structures of graphite by dissolved metal reduction in liquid ammonia.[29]

In terms of possible applications of graphane, it has very interesting characteristics for hydrogen storage. Its volumetric capacity of 0.12 kg $H_2$/L is higher than the DOE target of 0.081 kg $H_2$/L for the year 2015. Additionally, the gravimetric capacity of 7.7 wt. % H is higher than the 6 wt. % H DOE target for 2010. However, issues pertaining to the load and release kinetics will need to be explored. Alternatively, doping this purely two-dimensional semiconductor can be a source of a high mobility two-dimensional electron gas with variable concentration and have interesting electronics applications. In summary, it is fascinating that such a simple member of the family of hydrocarbons has been missing until now. If synthesized, it will not only be a noteworthy example of the predictive power of electronic structure methods but also will open a new window into the captivating world of low dimensional materials.

We want to thank Angela Lueking and Humberto R. Gutierrez for sharing their data with us and for many useful conversations. We have greatly benefited from conversations with Jayanth Banavar, Milton Cole, Vin Crespi, Peter Eklund, Jerry Mahan, Tom Mallouk, and Peter Schiffer. We also thank Ignacio Sofo for his help with the figures.